\documentclass[aps,prl,twocolumn,groupedaddress]{revtex4}
\usepackage{graphicx}

\begin{document}

\title{
Vibrational properties of  alpha- and sigma-phase Fe-Cr alloy
}

\author{S. M. Dubiel}
\email[Corresponding author: ]{dubiel@novell.ftj.agh.edu.pl}
\author{J. Cieslak}
\affiliation{Faculty of Physics and Applied Computer Science,
AGH University of Science and Technology, al. Mickiewicza 30, 30-059 Krakow, Poland}

\author{W. Sturhahn}
\affiliation{Advanced Photon Source, Argonne National Laboratory, 9700 South Cass Ave, Argonne
Illinois 60439, USA}

\author{M. Sternik}
\author{P. Piekarz}
\affiliation{Institute of Nuclear Physics, Polish Academy of Sciences, PL-31-342 Krakow, Poland}

\author{S. Stankov}
\affiliation{European Synchrotron Radiation Facility, BP 220, F-38043 Grenoble, France.
Present address: Karlsruher Institute of Technology, DE-76344 Eggenstein-Leopoldshafen,
Germany}

\author{K. Parlinski}
\affiliation{Institute of Nuclear Physics, Polish Academy of Sciences, PL-31-342 Krakow, Poland}

\date{\today}

\begin{abstract}
Experimental investigation as well as theoretical calculations, of the
Fe-partial phonon density-of-states (DOS) for nominally
Fe$_{52.5}$Cr$_{47.5}$ alloy having (a) $\alpha$- and (b) $\sigma$-phase
structure were carried out.  The former at sector 3-ID of the Advanced Photon
Source, using the method of nuclear resonant inelastic X-ray scattering, and
the latter with the direct method [K.  Parlinski et al., Phys.  Rev.  Lett.
{\bf 78}, 4063 (1997)].  The characteristic features of phonon DOS, which
differentiate one phase from the other, were revealed and successfully
reproduced by the theory.  Various data pertinent to the dynamics such as
Lamb-M\"ossbauer factor, $f$, kinetic energy per atom, $E_k$, and the mean
force constant, $D$, were directly derived from the experiment and the
theoretical calculations, while vibrational specific heat at constant volume,
$C_V$, and vibrational entropy, $S$ were calculated using the Fe-partial DOS.
Using the values of $f$ and $C_V$, we determined values for Debye temperatures,
$\Theta_D$.  An excellent agreement for some quantities derived from
experiment and first-principles theory, like $C_V$ and quite good one for
others like $D$ and $S$ was obtained.  \end{abstract}


\maketitle

For many years Fe-Cr alloy system has been of exceptional scientific and
technological interest.  The former stems on one hand from its interesting
physical properties such as magnetic ones, and on the other hand from the
fact that it forms a solid solution within the whole concentration range
preserving, at least metastably, the same crystallographic structure (bcc).
This, in turn, gives a unique chance for investigating the influence of the
composition on various physical properties within the same structure as well
as adequatly testing different theoretical models and theories.  The latter
follows from the fact that Fe-Cr alloys constitute the basic ingredient of
stainless steels (SS) that for a century have been one of the most
important structural materials \cite{Lo09}, and, consequently, some properties of SS
are inherited from the parent alloy.  To the latter belongs a $\sigma$-phase
that can precipitate if a quasi-equiatomic Fe-Cr alloy undergoes an
isothermal annealing in the temperature range $\sim800$K$\le$T$\le\sim1100$
K.  The $\sigma$-phase has a tetragonal structure (type D$^{14}_{4h}$
P$4_2$/mnm) with 30 atoms distributed over five different sites (Table 1).  Its
physical properties are, in general, quite different than those of the
$\alpha$-phase of similar composition.  Some properties, like the magnetic
ones, are even dramatically different \cite{Cieslak08a}, other properties, like the Debye
temperature, seem to be very similar \cite{Cieslak06}.  The latter is rather unexpected as
the hardness of the $\sigma$-phase is by a factor of $\sim 3$ larger than
that of the $\alpha$-phase.  To clarify this situation a more detailed
knowledge of vibrational properties of the $\sigma$- and $\alpha$-phases is
essential.  In addition, the $\sigma$-phase belongs to an important family of
tetrahedrally close-packed Frank-Kasper phases and is one of the closest
low-order crystalline approximants for dodecagonal quasicrystals which have
similar local structural properties with the icosahedral glass (ICG) \cite{vibrat00}.
The latter implies that the study of the vibrational properties of the
$\sigma$-phase should shed some light on similar properties in ICGs.
Challenged by this possibility and motivated by a lack of available knowledge
on dynamical properties of the real $\sigma$-phase we have carried out both,
an experimental investigation as well as theoretical calculations, of the
Fe-partial phonon density-of-states (DOS) for nominally
Fe$_{52.5}$Cr$_{47.5}$ alloy having (a) $\alpha$- and (b) $\sigma$-phase
structure.

The master alloy ($\alpha$-phase) was prepared by melting, in
appropriate proportion, $^{57}$Fe- enriched ($\sim$95\%) iron with chromium.
The ingot was then cold-rolled down to a thickness of about 30$\mu m$  from
which two $5\times5$ mm$^2$ plates were cut out.  They were next solution- treated
at 1273 K for 72 h.  One of the samples was afterwards transformed into the
$\sigma$-phase by the isothermal annealing at 973 K for 7 days.  The
verification of the transformation into the $\sigma$-phase caused by such
thermal procedure was done by recording a $^{57}$Fe M\"ossbauer spectrum at
295 K.  Experiments were conducted at sector 3-ID of the Advanced Photon
Source.  The vibrational properties such as the Fe-partial phonon
density-of-states were studied using the method of nuclear resonant inelastic
X-ray scattering (NRIXS) \cite{Sturhahn95,Sturhahn04}.  Synchrotron radiation X-rays were
monochromatized to a bandwidth of 1.2 meV and tuned in energy ranges of +/-
80 meV (room temperature measurements) and -20 meV to +80 meV (measurement at
20 K) around the $^{57}$Fe nuclear transition energy of 14.4125 keV.  The
X-ray flux and beam size at the sample position were $4\times10^9$ photons/s
and 0.3$\times$2 mm$^2$, respectively.  Data collection times were about 3
hours for the room temperature measurement of each sample and about 12 hours
for the low temperature measurement of the $\sigma$-phase sample.  We
followed previously described evaluation procedures \cite{Sturhahn95,Sturhahn04} using the publicly
available PHOENIX software \cite{Sturhahn00}.  The following quantities were derived
directly from the data:  Lamb-M\"ossbauer factor, $f$, kinetic energy per
atom, $E_k$, and the mean force constant, $D$.  No specific assumptions about
the character of the vibrations had to be made to obtain these values.  The
Fe-partial DOS was derived by direct data inversion using the Fourier-Log
method under the assumption of quasi- harmonic vibrations.  The consistency
of this procedure was verified by independent calculation of Lamb-M\"ossbauer
factor, kinetic energy per atom, and mean force constant from the DOS and by
agreement of these values with same quantities obtained directly from the
data.  Then the following quantities were calculated using the Fe-partial
DOS:  vibrational specific heat at constant volume, $C_V$, and vibrational
entropy, $S$.  The assignment of Debye temperatures, $\Theta_D$,
is based on the Debye
model, i.e., the DOS is proportional to energy squared, and they are widely
used in the literature.  With the determination of the Fe-partial DOS, we
have surpassed the Debye model but find it useful to provide Debye
temperatures for comparison.  Using the values of $f$ and $C_V$, we
determined commonly presented values for $\Theta_D$.

\begin{table}[t] 
\caption{\label{table1a} Atomic crystallographic positions and numbers of NN atoms for the five lattice sites of the Fe-Cr $\sigma$-phase.}
\begin{tabular}{|l|l|c|c|c|c|c|c|} \hline
Site& Crystallographic positions & \multicolumn{6}{c|}{ NN}                          \\ \hline
    &           & A   & B   & C   & D   & E  & Total  \\ \hline
A   & 2i (0,     0,     0    )        & -   & 4   & -   & 4   & 4  & 12              \\ \hline
B   & 4f (0.4,   0.4,   0    )        & 2   & 1   & 2   & 4   & 6  & 15              \\ \hline
C   & 8i (0.74,  0.66,  0    )        & -   & 1   & 5   & 4   & 4  & 14              \\ \hline
D   & 8i (0.464, 0.131, 0    )        & 1   & 2   & 4   & 1   & 4  & 12              \\ \hline
E   & 8j (0.183, 0.183, 0.252)        & 1   & 3   & 4   & 4   & 2  & 14              \\ \hline
\end{tabular}
\end{table}

In
calculations, both phases of Fe-Cr alloy were modeled by the appropriate
atomic configurations placed in a supercell with the periodic boundary
conditions.  The disordered $\alpha$-Fe$_{52.5}$Cr$_{47.5}$ alloy was
approximated by the $\alpha$-Fe$_{50}$Cr$_{50}$ one, for which we used the
$2\times 2\times2$ bcc cell with 16 atoms.  For random distribution of
atoms, there are about 500 different atomic configurations to be considered.
However, for the sake of computer and time capacity, we have chosen at random
only five to be included in our calculations.  The real
$\sigma$-Fe$_{52.5}$Cr$_{47.5}$ sample was approximated by a
$\sigma$-Fe$_{53.3}$Cr$_{46.7}$ one.  The latter was studied in the
$1\times1\times1$ tetragonal supercell with 30 atoms (16 Fe and 14 Cr atoms).  The
structure optimization was done using the spin-polarized density functional
total energy calculations performed within the generalized gradient
approximation (GGA) using the VASP package \cite{Kresse08a,Kresse08b}.  The valence electrons for
each atom (electron configuration:  $d^5s^1$ and $d^6s^2$ for the Cr and Fe
atoms, respectively) are represented by plane wave expansions.  The wave
functions in the core region are evaluated using the full-potential projector
augmented-wave (PAW) method \cite{Blochl94,Kresse99}.  The integrations in the reciprocal space
were performed on the $8\times8\times8$ and $4\times4\times4$ grid for
$\alpha$- and $\sigma$-phase, respectively.  During the optimization, the
Hellmann-Feynman (H-F) forces and the stress tensor were calculated and the
structure optimization was performed in two steps.  First the lattice
constants were determined assuming the appropriate symmetry, then the atomic
positions were found in a fixed unit cell.  The crystal structure
optimization was finished when residual forces were less than $10^{-5}$ eV/\AA
and stresses were less than 0.1 kbar.  The calculated lattice constants are
$a=5.64$\AA and $a=8.60$\AA, $c=4.74$\AA for the cubic and tetragonal
symmetry, respectively.
The optimized magnetic moments on the Fe atoms
are ordered ferromagnetically with the average
values 2.03 and 0.96  $\mu_B$, in the $\alpha$
and $\sigma$ phase, respectively.
On the Cr atoms, the antiparallel arrangement occurs
with the negative mean magnetic moments
-0.17 and -0.28  $\mu_B$ in the respective phases.

For the optimized structures the phonon dispersions
and density of states were calculated using the direct method \cite{Parlinski97,Parlinski07}.  The
dynamical matrix of the crystal is constructed from the H-F forces generated
while displacing atoms from their equilibrium positions.  For considered
structures each atom must be displaced in three directions.  For the $\alpha$
and $\sigma$-phases, a complete set of H-F forces is obtained from 48 and 90
independent atomic displacements, respectively.  The amplitude of the
displacements equals 0.03 \AA.  To minimize systematic errors we applied
displacements in positive and negative directions.  Finally, the phonon
frequencies are obtained by the diagonalization of the dynamical matrix for
each wave vector.  The phonon DOS is calculated by the random sampling on the
k-point grid in the reciprocal space, and then the thermodynamic functions
are obtained within the harmonic approximation.

\begin{figure}[hbt]
\includegraphics[width=.49\textwidth]{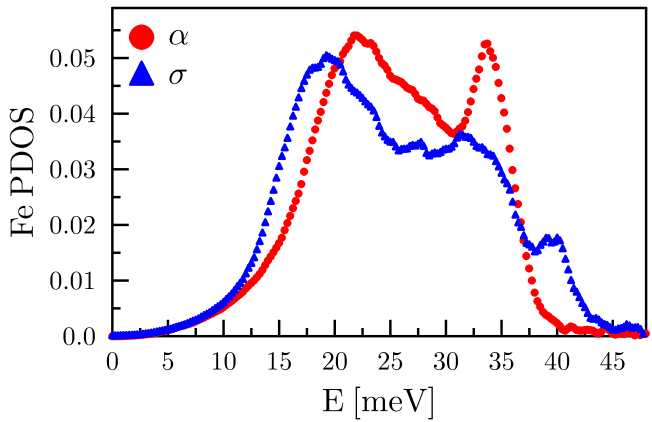}
\caption{(Color online) Phonon DOS as measured on $^{57}$Fe atoms at 298 K for the
$\alpha$- (circles) and the $\sigma$-phase (triangles) on the Fe$_{52.5}$Cr$_{47.5}$ samples. }
\label{fig1}
\end{figure}

The phonon DOS's measured on $^{57}$Fe for both phases of the Fe-Cr alloy are
presented in Figure 1.  The differences in the energy range covered by the
spectrum and the discrete structure of the spectrum are significant.  The
spectrum obtained for the $\alpha$-phase is found to be similar to that of
pure Fe \cite{Lucas08} exibiting a distinct peak at 36 meV.  The Fe-partial DOS
spectrum of the $\sigma$-phase demonstrates the additional high-frequency
peak at 40 meV, that is not observed neither in $\alpha$-FeCr nor in pure
bcc-Fe.  There is also the shift downward of the low-energy peak, which makes
the entire spectrum broader than in the $\alpha$-phase.  In Figure 2 the
measured and calculated phonon DOS spectra of $\alpha$-FeCr are compared.
The shape of both spectra is similar and two characteristic peaks of measured
spectrum are reproduced satisfactorily.  As in the chosen supercell, there
are eight Fe atoms, hence by taking into account five different atomic
configurations, one considers vibrations of forty independent Fe atoms.
Their partial DOS turned out to be different, but their shape was not
correlated with the particular nearest-neighbor (NN) - next NN shell.
Consequently, the final DOS was calculated using the partial contributions
with the same weights.  The discrepencies between measured and calculated
spectrum are likely caused by an incomplete representation of possible atomic
configuration of FeCr disordered alloy in our model.

\begin{figure}[hbt]
\includegraphics[width=.49\textwidth]{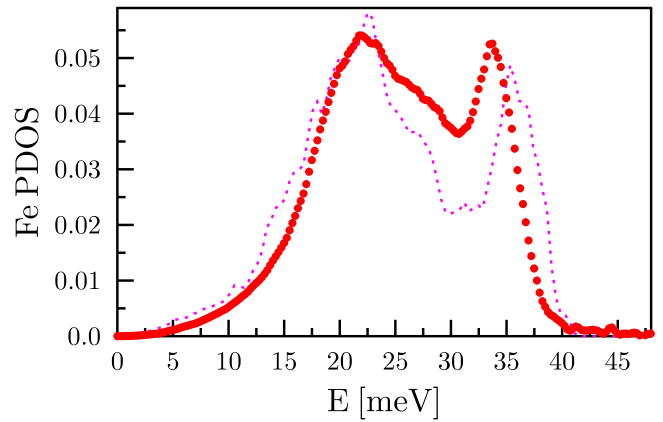}
\caption{(Color online) Phonon DOS as measured on $^{57}$Fe atoms at 298 K
(circles) and as calculated (dotted line) for the $\alpha$-Fe$_{52.5}$Cr$_{47.5}$. }
\label{fig2}
\end{figure}

Likewise, the calculations of $\sigma$-FeCr performed for only one
configuration yield the phonon DOS exhibiting characteristic features of the
experimental spectrum (Figure 3).  Observed discrepancies, like
underestimation of the intensity of low-frequency peak or shift of
high-frequency peak, are not significant.  Using the theoretical result, it
is feasible to separate the individual contributions to the total DOS
generated by the Fe atom placed at each particular crystallographic position.

Thus, we see that the Fe atoms on sites A and C are
causing the high-energy contributions to the DOS.The NN sites of these Fe atoms
are placed at distances shorter than 2.48\AA.  In pure bcc-Fe, all 8 NN
atoms are situated at the same distance of 2.485\AA.  The shorter
distances between atoms result in the larger interatomic interation.
Therefore, in $\sigma$-FeCr, the phonon frequencies higher than 40 meV, which
is a high-frequency limit of bcc-Fe DOS spectrum, are observed.  Also due to
larger dispersion of distances in the $\sigma$-phase, its phonon spectrum is
much broader than in the $\alpha$-phase, where the atomic positions are very
close to those of the pure bcc-Fe.

\begin{figure}[hbt]
\includegraphics[width=.49\textwidth]{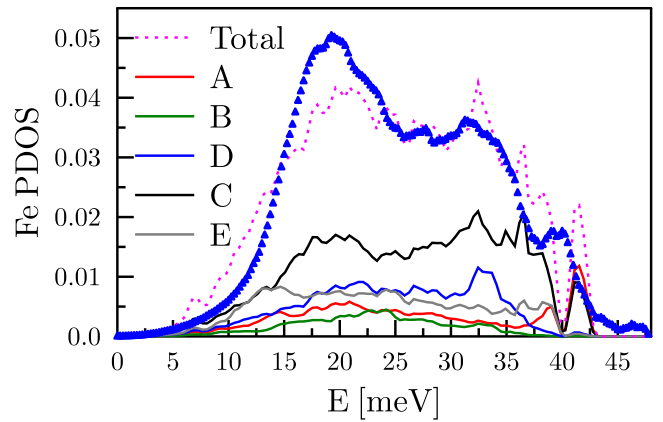}
\caption{(Color online) Phonon DOS as measured on $^{57}$Fe atoms at 298 K on the
$\sigma$- Fe$_{52.5}$Cr$_{47.5}$ sample (triangles) and as calculated (dotted line).
DOS-curves for particular crystallographic sites are indicated, too. }
\label{fig3}
\end{figure}

The data derived from the experiment and the theoretical calculations are
displayed in Table I.  One can see a very good agreement for some quantities
like specific heat and quite good for other ones like force constant and
entropy.  The values of the Debye temperature derived from the specific heat
are close to those calculated from the Lamb-M\"ossbauer factor, and there is
a small difference for this quantity found for different phases.  These
findings disagree with experimentally found $\Theta_D$-values for the two
phases using second-order Doppler shifts from M\"ossbauer spectroscopy \cite{Cieslak06}.
In those studies, the $\Theta_D$-values for the $\alpha$-phase were larger
than the ones for the $\sigma$-phase.  This apparent discrepancy is explained
by the fact that the center shift measured by M\"ossbauer spectroscopy
constitutes a sum of the chemical isomer shift, which is independent on the
atomic motion, and the second-order Doppler shift, which is a relativistic
correction to the atomic energy levels purely due to motion.  The
second-order Doppler shift is proportional to the vibrational kinetic energy
of the $^{57}$Fe atom \cite{Sturhahn99}, i.e. SOD[mm/s] = -0.00565$\times E_k$[meV].
Only under the assumption that the chemical isomer shift is temperature
independent, the M\"ossbauer measurement can provide the correct Debye
temperature.  This does not seem to be the case here, and our NRIXS data
present a notable improvement in the understanding of the role of vibrations
in the Fe-Cr system.  The value of the vibrational entropy measured at 298 K
for the $\alpha$-phase agrees quite well with inelastic neutron scattering
results on samples of similar composition \cite{Lucas08}.  Also the difference in the
entropy values, $\Delta S = S_\sigma-S_\alpha=0.095\pm 0.009k_B$ as determined
in the present experiment from the Fe-partial DOS agrees well with the
corresponding difference calculated from the equation $\Delta S = 3 k_B
\ln(\Theta_{D\sigma}/\Theta_{D\alpha}) = 0.07 k_B$, where $k_B$ is the
Boltzmann constant and $\Theta_{Di}$ is the Debye temperature as determined
from the second-order Doppler shift for the $\alpha$ ($i=\alpha$) and the
$\sigma$ ($i=\sigma$) phase, respectively \cite{Cieslak06}.  The corresponding theoretical
value of $\Delta S$ is equal to 0.058 $k_B$.  Taking into account the
approximations used in the calculations, the agreement seems to be quite
satisfactory.

\begin{table}
\caption{\label{table} Physical quantities derived from the measured and calculated
Fe-partial DOS.  The units are meV/atom for vibrational kinetic energy,
$E_k$; $k_B$/atom for vibrational entropy, $S$, and specific heat $C_V$; N/m
for mean force constant, $D$; K for Debye temperatures, $\Theta_D$.}
\begin{tabular}{|l|l|l|l|l|} \hline
         &\multicolumn{2}{c|}{experiment}& \multicolumn{2}{c|}{theory}                   \\ \hline
quantity             & $\alpha$        & $\sigma$         & $\alpha$ & $\sigma$         \\ \hline
$f$@298K             & 0.782$\pm0.001$ & 0.768$\pm0.002$  &          &                  \\ \hline
$E_k$@298K           & 42.4$\pm0.1$    & 41.4$\pm0.2$     &          &                  \\ \hline
$D$@298K             & 156$\pm1$       & 157$\pm2$        & 149.41   & 150.59           \\ \hline
$C_V$@298K           & 2.747$\pm0.006$ & 2.748$\pm0.007$  & 2.746    & 2.752            \\ \hline
$S$@298K             & 3.252$\pm0.006$ & 3.347$\pm0.007$  & 3.354    & 3.412            \\ \hline
$f$@20K              &                 &0.9150$\pm0.0002$ &          &                  \\ \hline
$E_k$@20K            &                 & 18.93$\pm0.06$   &          &                  \\ \hline
$D$@20K              &                 & 155.1$\pm0.7$    &          &                  \\ \hline
$f$@0K               &0.9194$\pm0.0001$&0.9149$\pm0.0003$ &          &                  \\ \hline
$E_k$@0K             & 19.32$\pm0.07$  & 18.95$\pm0.06$   &          &                  \\ \hline
$\Theta_D$($f$@298K) & 417             & 398              &          &                  \\ \hline
$\Theta_D$($C_V$)    & 399             & 398              &          &                  \\ \hline
$\Theta_D$($f$@20K)  &                 & 387              &          &                  \\ \hline
$\Theta_D$($f$@0K)   & 403             & 385              &          &                  \\ \hline
\end{tabular}
\end{table}

In summary, we have revealed, both experimentally and theoretically, significant differences in
the partial-Fe phonon DOS of the $\alpha$ and $\sigma$
phases of a quasi-equiatomic Fe-Cr alloy.
Although, the sigma phase is a very complex object
due to a high number of atoms per unit cell,
five different sublattices with high coordination numbers (12-15), each showing
chemical disorder, which altogether results in a huge number of possible
atomic configurations, it was described reasonable well
in terms of only one adequately chosen configuration.
From the calculations, it is also evident that the dynamics in particular
sublattices is different. The method was also successfully used to calculate
the dynamics of the disordered alloy in the alpha-phase.
Although its crystallographic structure is much simpler, but, due to the
chemical disorder, the number of possible atomic configurations is high,
making the calculations not trivial. We have also obtained relevant
thermodynamic quantities without necessity of using empirical parameters.
Such a complex alloy has been studied for the first time within
the combined NRIXS and theoretical {\it ab initio} approach,
and it may provide understanding of lattice dynamics in
a wide variety of disordered systems.

\begin{acknowledgments}
The results reported in this study were partly obtained
within the project supported by the Ministry of Science and Higher Education,
Warsaw (grant No. N N202 228837) and the Project no.  44/N-COST/2007/0.
Use of the Advanced Photon Source was supported by the
U. S. Department of Energy, Office of Science, Office of
Basic Energy Sciences, under Contract No. DE-AC02-06CH11357.
\end{acknowledgments}

\end{document}